\documentclass[journal=jpccck,manuscript=article, layout=twocolumn]{achemso}
\setkeys{acs}{usetitle = true}

\usepackage[version=3]{mhchem} 
\usepackage{graphicx}
\usepackage{color}
\usepackage{soul}
\usepackage{hyperref}
\usepackage{multirow}

\newcommand{\onlinecite}[1]{\hspace{-1 ex} \nocite{#1}\citenum{#1}} 

\author{Zhiming Wang}
\author{Xianfeng Hao}
\author{Stefan Gerhold}
\author{Petr Mares}
\author{ Margareta Wagner}
\author{Roland Bliem}
\affiliation{Institute of Applied Physics, Vienna University of Technology, Vienna, Austria.}
\author{Karina Schulte}
\affiliation  
{MAX IV Laboratory, Lund University, Lund, Sweden}
\author{Michael Schmid}
\affiliation{Institute of Applied Physics, Vienna University of Technology, Vienna, Austria.}
\author{Cesare Franchini}
\affiliation{Faculty of Physics and Center for Computational Materials Science, University of Vienna, Vienna, Austria.}
\author{Ulrike Diebold}
\affiliation{Institute of Applied Physics, Vienna University of Technology, Vienna, Austria.}

\title[\texttt{achemso} demonstration]
{Stabilizing Single Ni adatoms on a Two-dimensional Porous Titania Overlayer at the SrTiO$_3$(110) Surface}

\begin{document}
\begin{abstract} 
Nickel vapor deposited on the SrTiO$_3$(110) surface was studied using scanning tunneling microscopy (STM), photoemission spectroscopy (PES) and density functional theory (DFT) calculations. This surface forms a (4~$\times$~1) reconstruction, comprised of a two-dimensional titania structure with periodic six- and ten-membered nanopores. Anchored at these nanopores, Ni single adatoms are stabilized at room temperture. PES measurements show that the Ni adatoms create an in-gap state located at 1.9\,eV below the conduction band minimum and induce an upward band bending. Both experimental and theoretical results suggest that Ni adatoms are positively charged. Our study produces well-dispersed single adatom arrays on a well characterized oxide support, providing a model system to investigate single adatom catalytic and magnetic properties.
\end{abstract}


\section{Introduction}
Scanning probe microscopy studies of single adatoms on surfaces have revealed novel physical phenomena \cite{Ternes09jpcm,Wiesendanger11cossms}. In addition, single metal adatoms on oxide supports have shown remarkable performance in catalytic reactions \cite{Yang:ACR13, Lin:natchem11, Lin:jacs13, Moses:jacs13}.  Approaches to produce single metal adatom arrays include mass-selected soft-landing, wet-chemistry approaches, as well as vapor deposition in ultra-high vacuum conditions \cite{Kaden:sc09,Yang:ACR13,Novotny:prl12}. However, stabilizing single atoms on oxide supports has remained a significant challenge because sintering occurs under realistic reaction conditions \cite{Giordano:prl08, Nilius:ssr09, Parkinson:natm13}. Special sites such as defects, moir\'{e} patterns and reconstructions with strong modulation of surface potential \cite{Novotny:prl12} enable to anchor and stabilize single metal adatoms. Recently, Freund and co-workers have demonstrated that a two-dimensional (2D) porous silica structure grown on metal substrates is a suitable candidate for stabilizing single adatoms such as Li, Fe, Ag and Pd  \cite{Martinez:prl09,Jerratsch:acsnano10,Ulrich:ss09}. This silica structure is comprised of a single layer of corner-sharing SiO$_4$ tetrahedra that form six-membered rings of 5\,{\AA} diameter \cite{Weissenrieder:prl05}. Here we introduce a 2D porous titania structure on SrTiO$_3$ that can serve as a template for single metal adatoms. 

Strontium titanate (SrTiO$_3$), the prototypical perovskite oxide, has attracted extensive interest \cite{Ohtomo:nat04,Kan:05natm,OHTA:natm07,Kudo:09csr,Santander:nat11,Wang:pnas14}. Surface reconstructions on various SrTiO$_3$ faces often consist of porous 2D titania structures \cite{Kienzle:prl11, Enterkin:natm10, Marks:ss09, Biswas:apl11}. For the (110) orientation, a 2D titania overlayer consisting of a single layer of TiO$_4$ tetrahedra resides directly on the last (SrTiO)$^{4+}$ plane \cite{Enterkin:natm10}. The tetrahedra share oxygen corners and form networks of rings of variable sizes. For example, six- and ten-membered rings are found on the SrTiO$_3$(110)-(4~$\times$~1) surface [see Figure\,\ref{Fig1}(b)] \cite{Enterkin:natm10}. The six-membered ring has a diameter of 5.5\,{\AA}, providing a perfect site for accommodating single Sr adatoms \cite{Wang:prl13}. The Sr adatoms [Figure\ \ref{Fig1}(a)] are an integral part of the  structure \cite{Wang:prl13}, as they assist in compensating the polarity inherent in the (110) surface. The Sr adatoms are well-dispersed and have remarkably high thermal stability. They thus could serve as nucleation centers and guide the growth of an array of noble metal nanostructures \cite{Zhang:jcp11}. In this Letter, we explore the formation of Ni adatoms on the 2D porous titania structure on the SrTiO$_3$ surface. Ni/SrTiO$_3$ can be considered as a model system to investigate single atom catalysis in, for example, water splitting \cite{Townsend:12ees}. Scanning tunneling microscopy (STM) measurements show that two types of single Ni adatoms adsorb at the six- and ten-membered rings, respectively. Photoemission spectroscopy (PES) experiments show that the Ni adatoms introduce an in-gap state and an upward band bending. Experimental and theoretical results suggest that the Ni adatoms are positively charged.

\section{Methods}
STM experiments were performed in an ultra high vacuum (UHV) system with a SPECS Aarhus STM at room temperature (RT) \cite{Wang:jpcc13}. Synchrotron radiation PES experiments were performed at beamline I311 at the MAX IV Laboratory \cite{Nyholm:nucl01}. The pressure in both UHV systems was better than 1$\times$10$^{-10}$\,mbar. Nb-doped (0.5\,wt$\%$) SrTiO$_3$ single crystals (5\,mm\,$\times$\,5\,mm\,$\times$\,0.5\,mm) were purchased from CrysTec, Germany. A clean surface was prepared by cycles of Ar$^+$ sputtering (1\,keV, 5\,$\mu$A, 10\,minutes) followed by annealing in O$_2$ at pressures of 2$\times$10$^{-6}$\,mbar at 1000\,$^{\circ}$C for 1\,h \cite{Wang:apl09}. The samples were heated by electron bombardment (13\,mA, 900\,V) at the back and the temperature was monitored with an infrared pyrometer. High purity (99.999\%) Ni metal was deposited on the surface at RT by an e-beam evaporator (Omicron EFM3). The deposition rate of 0.05\,\AA/min was calibrated using a quartz crystal microbalance. DFT calculations were carried out with the ``Vienna ab initio simulation package'' (VASP) code \cite{vasp1, vasp2}. We adopted the projector augmented-wave method  \cite{Blochl:prb94} and the Perdew-Burke-Ernzerhof functional  \cite{pbe} with a kinetic energy cutoff of 600 eV for plane waves. A Monkhorst-Pack $k$-$point$ mesh (2\,$\times$\,3\,$\times$\,1) was used. The surface structure was modeled with a supercell that was symmetric along the [110] direction, and consisted of a 9-layer slab separated by a vacuum layer of 12\,\AA{}. The atoms in the central three layers were fixed, and the other atoms were allowed to relax until the force on each atom was less than 0.02\,eV/\AA{}. Simulated STM images were generated with the Tersoff-Hamann approximation \cite{Tersoff:prl83} by integrating the local density of states from the Fermi level to 1.5\,eV above the conduction band edge. In order to take electronic correlation into accout,  an additional on-site Coulomb repulsion term with $U_\mathrm{eff}$=4.5/5.5\,eV was applied to the Ti/Ni 3$d$ states, respectively.

\section{Results}

\begin {figure}
\includegraphics [width=3.4 in,clip] {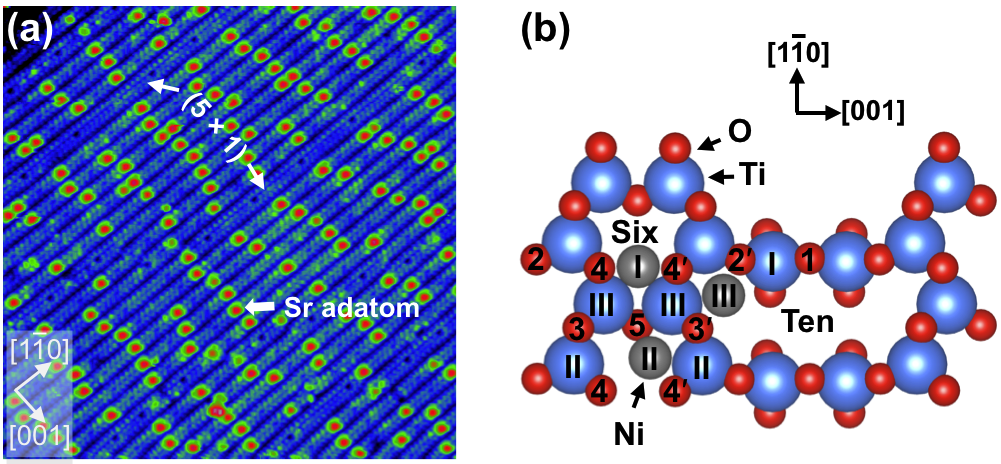}
\caption{
(color online). (a) STM image (30~$\times$~30 nm$^2$, V$_{sample}$ = +2.0 V, I$_{tunnel}$ = 0.3 nA) of the SrTiO$_3$(110)-(4~$\times$~1) surface. Labeled are Sr adatoms as well as a few stripes that have a (5~$\times$~1) periodicity.\cite{Wang:prl13} (b) Top view of the SrTiO$_3$(110)-(4~$\times$~1) surface. Ti and O atoms are shown in blue and red, respectively. Positions (I-III) for Ni adatoms (gray), attached to surface O atoms, have the most favorable  adsorption energies according to DFT calculations.  
}
\label{Fig1}
\end{figure}

\begin {figure}
 \includegraphics [width=3.4 in,clip] {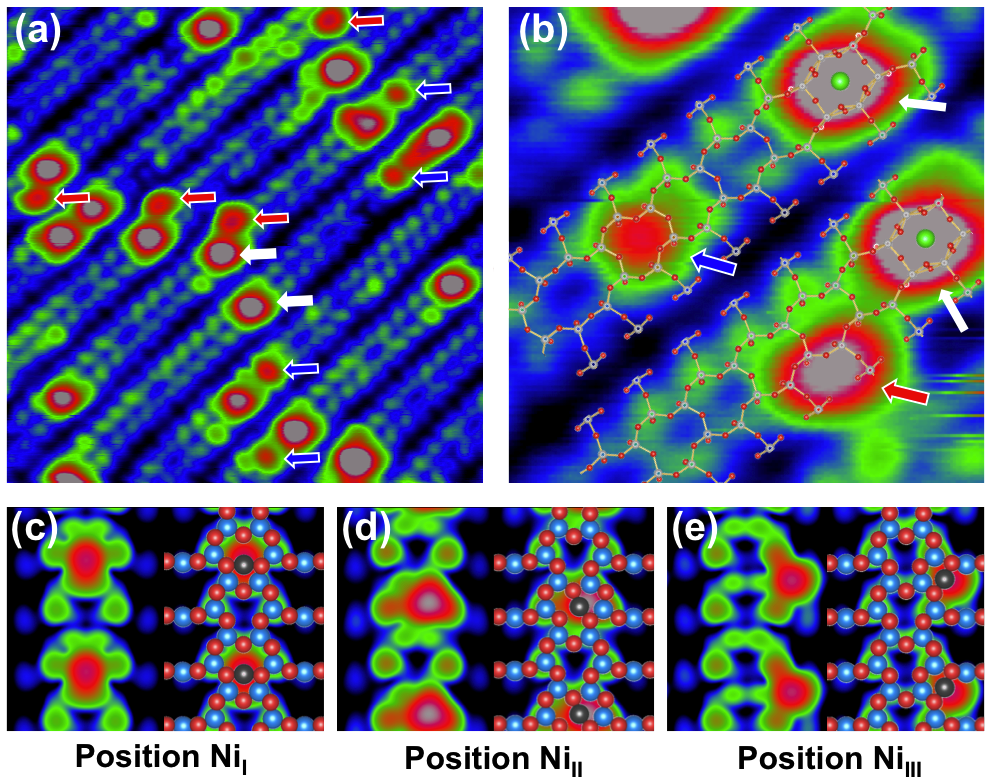}
 \caption{
(color online). (a) STM image (10~$\times$~10\,nm$^2$, V$_{sample}$ = +2.0\,V, I$_{tunnel}$ = 0.3\,nA) of 0.01\,{\AA} Ni deposited on the SrTiO$_3$(110) surface at RT. Marked with arrows are Sr atoms (white) and Ni atoms at the center (blue) and side (red) of the rows. (b) High resolution STM image (3.2~$\times$~3.2\,nm$^2$, V$_{sample}$ = +2.0\,V, I$_{tunnel}$ = 0.3\,nA) with a structural model superimposed. (c-e) DFT simulated STM images of single Ni adatoms adsorbed at the center (c) and off-center (d) of the rows (in a six-membered ring), and at the side (e) of the row (in a ten-membered ring).
}
\label{Fig2}
\end{figure}

STM images of the SrTiO$_3$(110)-(4~$\times$~1) surface exhibit quasi-one-dimensional stripes along the [1$\overline{1}$0] direction [see Figure\ \ref{Fig1}(a)] \cite{Wang:prb11}. Each (4~$\times$~1) stripe contains two bright rows of periodic dots, corresponding to the Ti$_\mathrm{III}$ and Ti$_\mathrm{II}$ atoms in the six-membered rings  [see Figure\ \ref{Fig1}(b)] \cite{Li:prl11, Wang:prl13}. The stripes are separated by a dark trench originating from the Ti$_\mathrm{I}$ atoms in the ten-membered rings. Single Sr adatoms of typically apparent height of 240 pm are located in the middle of the six-membered rings, bonded to four oxygen atoms [see Figure\ \ref{Fig1}(b) and \ref{Fig2}(b)] \cite{Wang:prl13}.

In Figure\ \ref{Fig2}(a) we present an empty-state STM image following deposition of 0.01\,{\AA} Ni on the SrTiO$_3$(110) surface at RT. In addition to Sr adatoms, two types of bright protrusions are observed. Each of these protrusions has the same adsorption site and an apparent height of $\sim$\,150\,pm, thus we conclude each bright protrusion contains only one Ni atom. One type of Ni adatoms is located at the side of the (4~$\times$~1) stripes [labeled with red arrows in Figure\ \ref{Fig2}(a)]; the other one appears at the center of the stripes, similar to the Sr adatoms but with smaller size  [labeled with blue arrows in Figure\ \ref{Fig2}(a)]. Both types of Ni adatoms prefer to adsorb close to the intrinsic Sr adatoms. By superimposing a structural model on a high-resolution STM image it is apparent that the center Ni adatom is located in a six-membered ring, whereas the side Ni adatom is located at the corner of a ten-membered ring [see the red and blue arrows in Figure\ \ref{Fig2}(b)].   

In order to determine the adsorption sites and energies, we have performed DFT calculations of Ni adatoms at various sites of the SrTiO$_3$(110)-(4~$\times$~1) surface [see Figure\ \ref{Fig1}(b)]. A Ni atom (Ni$_{\mathrm{I}}$) attached between O4 and O4$^{'}$ atoms in the six-membered ring constitutes the most favorable configuration with an adsorption energy of 1.1 (-3.4)\,eV with the reference to a Ni atom in the bulk fcc lattice (gas phase) \cite{note} [see Figure\ \ref{Fig1}(b) and Table\ \ref{Table1}]. The adsorption energy is about 0.3\,eV less favorable when the Ni atom is placed between O4$^{'}$ and O5 atoms in the six-membered ring (Ni$_{\mathrm{II}}$), or between O2$^{'}$ and O3$^{'}$ atoms in the ten-membered ring (Ni$_{\mathrm{III}}$) [see Figure\ \ref{Fig1}(b)]. All other configurations are energetically less favorable, with a more than 1\,eV higher adsorption energy (not shown here). Note the clear dependence between bond length and adsorption energy, \textit{e.g.}, the shorter the bond length, the larger the adsorption energy.
In simulated STM images the Ni adatoms are present as bright protrusions in Figure\ \ref{Fig2}(c)-(e). We conclude that the Ni$_{\mathrm{I}}$ and Ni$_{\mathrm{III}}$ adatom observed in Figure\ \ref{Fig2}(b) reside in center positions (in 6-membered rings) and side positions (in 10-membered rings), respectively.

\begin {figure}[tb]
 \includegraphics [width=3.4 in,clip] {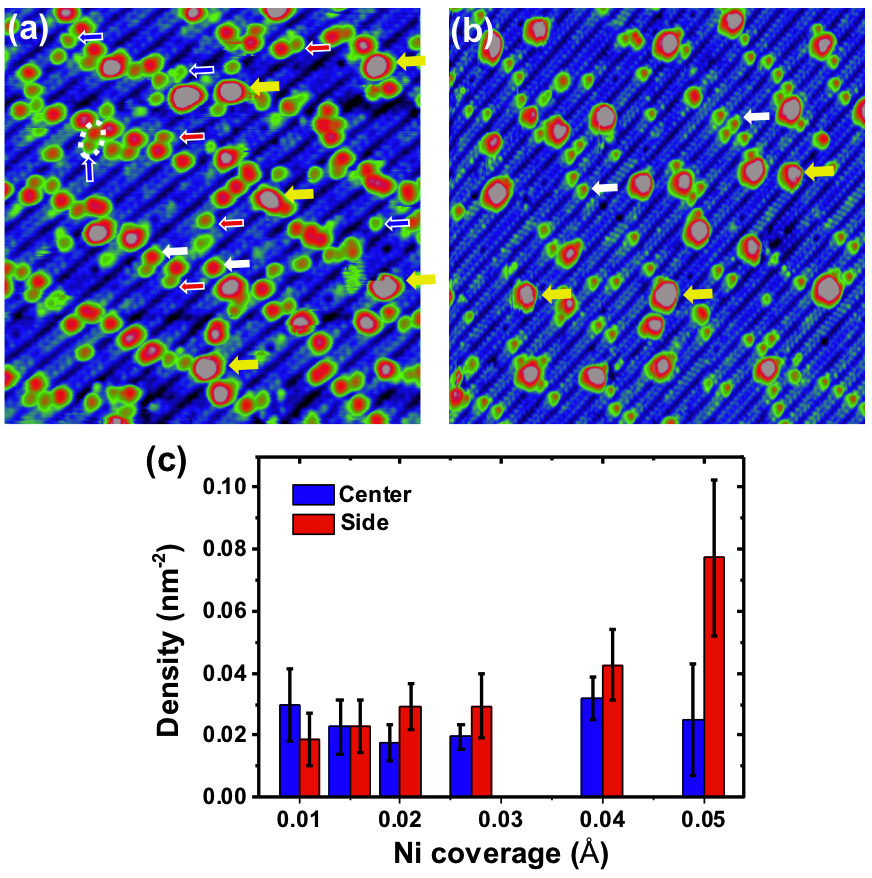}
 \caption{
(color online). (a) STM image (20~$\times$~20\,nm$^2$, V$_{sample}$ = +1.6\,V, I$_{tunnel}$ = 0.4\,nA) of 0.05\,{\AA} Ni deposited on the SrTiO$_3$(110) surface at RT. Marked with arrows are Sr atoms (white), Ni atoms at the center (blue) and side (red) of the rows and clusters (yellow). (b) STM image (30~$\times$~30\,nm$^2$, V$_{sample}$ = +2.0\,V, I$_{tunnel}$ = 0.3\,nA) after mild annealing 0.1\,{\AA} Ni deposited on the surface at RT. Marked with arrows are Sr atoms (white) and Ni clusters (yellow).(c) Histogram of the density of Ni adatoms adsorbed at center and side rows of the SrTiO$_3$(110)-(4~$\times$~1) surface for various Ni coverages. 
}
\label{Fig3}
\end{figure}

Figure\ \ref{Fig3}(a) shows a STM image after deposition of 0.05\,{\AA} Ni at RT. Ni single adatoms are again present and adsorbed near the intrinsic Sr adatoms, forming well-dispersed arrays. Statistics over a number of STM images [Figure\,\ref{Fig3}(c)] show that Ni adatoms prefer the center position (0.029\,$\pm$\,0.011\,nm$^{-2}$) to the side position (0.018\,$\pm$\,0.008\,nm$^{-2}$) at high coverage. As the coverage increases, the density of the side Ni adatoms increases to 0.077\,$\pm$\,0.025\,nm$^{-2}$, while the density of the center Ni adatom almost stays constant  (0.025\,$\pm$\,0.018\,nm$^{-2}$). In addition, clusters with apparent heights ranging from 300 to 400\,pm start to form. These clusters appears along the Sr meandering lines, attributing to Ni atoms adsorbing on the Sr adatoms [labeled with yellow arrows in Figure\,\ref{Fig3}(a)].

\begin {figure}[tb]
 \includegraphics [width=3.4 in,clip] {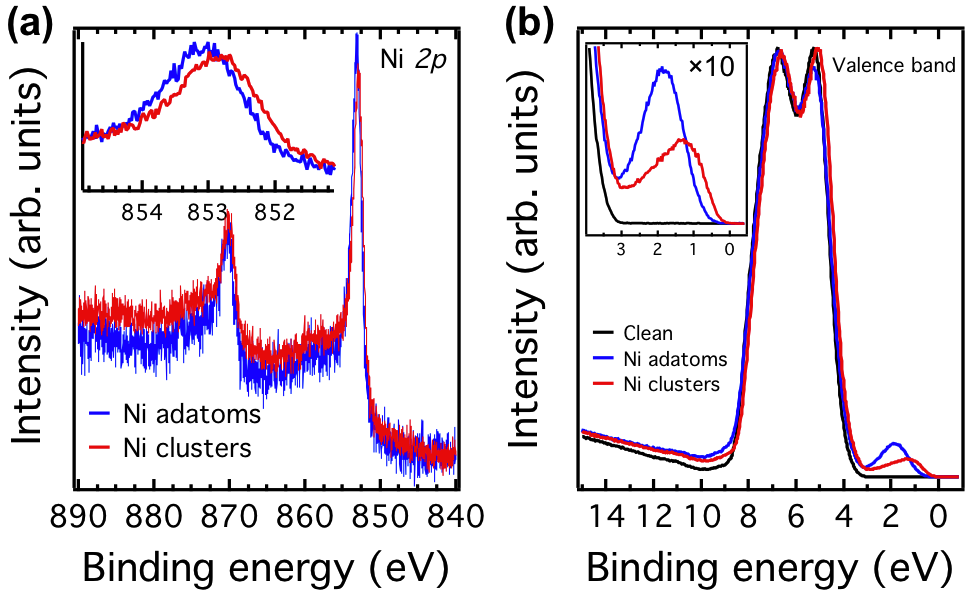}
 \caption{
(color online). (a) Ni $2p$ core-level photoemission spectra of  0.1\,{\AA} Ni adatoms (blue) and clusters (red) on the SrTiO$_3$(110)-(4~$\times$~1) surface. The inset shows a shift of the spectra by 0.2 eV. (b) Valence band photoemission spectra of the clean surface (black), surface with 0.1\,{\AA} Ni adatoms (blue) and Ni clusters (red). The inset shows surface states located at 1.9 and 1.3 eV below $E_{\mathrm{F}}$ in the gap region for adatoms and clusters, respectively. The core-level and valence band spectra were measured with photon energies of 1000 and 65 eV, respectively. All spectra were taken at RT.
}
\label{Fig4}
\end{figure}

\begin{table*}
\caption{Characterization of adsorption configurations of Ni adatom on the SrTiO$_3$(110)-(4~$\times$~1) surface. Listed are adsorption energy $E_\mathrm{ads}$ (eV/Ni atom)\cite{note}, the length of the Ni -- O bond (\AA), the O -- Ni -- O bonding angle \,($^{\circ}$), the height of the Ni atom (\AA) compared to the surface plane, magnetic moment ($\mu_\mathrm{B}$) as well 
as the Bader charge analysis for Ni adatoms in the three configurations shown in Figure\ \ref{Fig1}. The calculations were done within the GGA+U scheme.}
\begin{tabular}{cccc}
\hline\hline
   Configurations      & I       & II           & III   \\
\hline\hline
$E_\mathrm{ads}^\mathrm{bulk}$     &  1.11   &  1.49  &  1.38   \\
$E_\mathrm{ads}^\mathrm{gas}$  & -3.38 & -3.0 & -3.1 \\
Ni\,--\,O bonding length        & 1.790,1.790    &  1.828,1.839 &   1.819,1.832   \\
Bonding angle            & 172 &  162   &   168 \\
Height       & 0.801   &  0.468 & 0.596   \\
magnetic moments ($\mu_\mathrm{B}$)        & 0.484 & 0.927 &  0.21  \\
Bader charge  & +0.30 & +0.60 &  +0.20  \\
\hline\hline
\end{tabular}
\label{Table1}
\end{table*}

Figure\ \ref{Fig4}(a) shows the Ni $2p$ core-level photoemission spectra of a coverage of 0.1\,{\AA} Ni adatoms and clusters on the SrTiO$_3$(110)-(4~$\times$~1) surface. For the surface with Ni clusters, the Ni $2p_\mathrm{3/2}$ peak is positioned at 852.8 eV, for the adatoms it is shifted by 0.2 eV to a higher binding energy of 853.0 eV [see inset of Figure\ \ref{Fig4}(a)]. 

To further characterize the electronic structure, valence band photoemission spectroscopy measurements were performed [see Figure\ \ref{Fig4}(b)]. The valence band of the clean surface shows mainly O $2p$-derived features. The valence band maximum is located at 3.2 eV below the Fermi level and no surface states are observed in the band gap region [see inset in Figure\ \ref{Fig4}(b)] \cite{Wang:jpcc13,Wang:pnas14}. After depositing Ni adatoms, the whole spectrum shifts slightly to lower binding energy and an in-gap state with a binding energy of 1.9 eV is observed [see the blue curve in the inset of Figure\ \ref{Fig4}(b)]. On the surface with Ni clusters an in-gap state appears at a binding energy of 1.3 eV.

\section{Discussion}

The DFT results in Table\,\ref{Table1} predict that the center Ni adatom is more favorable than the side configuration, which is in accord with the STM results for Ni low coverages. Two side positions can be occupied in a (4~$\times$~1) unit cell, while just one position is available for the center Ni$_{\mathrm{I}}$ adatom [see the structural model in Figure\ \ref{Fig1}(b)]. This two-fold side adsorption position can simply explain the experimental observation of the higher density of side Ni adatom when increasing the Ni coverage [see Figure\, \ref{Fig3}(c)] .

Note that the adsorption energy for all adatom configurations is positive with respect to a Ni atom in bulk fcc lattice [see Table\ \ref{Table1}]. This implies that it is thermodynamically more favorable for Ni to form clusters on the SrTiO$_3$(110) surface. This is consistent with the experimental results that Ni adatoms can change into clusters (with apparent heights of $\sim$\,600\,pm) upon mild annealing (below 300\,$^{\circ}$C) in UHV [see Figure\ \ref{Fig3}(b)]. However, single Ni adatoms are preferred to form on the surface even at RT. On the one hand, Ni vapor will adsorb on the surface as single adatoms first, since the adsorption energy of the single Ni adatom on the surface is negative with respect to a Ni atom in the gas phase [see Table\ \ref{Table1}]. On the other hand, sintering of Ni adatoms is kinetically hindered on the surface. This indicates that the nanopores on the SrTiO$_3$-(4~$\times$~1) surface play an important role for anchoring and stabilizing single adatoms.

Although the origin of in-gap states on SrTiO$_3$ is still under debate \cite{Fujimori:jpcs96,Ishida:prl08}, these states typically appears at a binding energy of 1.3 eV, for example, through electron doping with atomic hydrogen or oxygen vacancies \cite{D'Angelo:prl12,Wang:jpcc13}. Note that the formation of in-gap states is accompanied with a downward surface band bending due to electrons confined in the near surface region \cite{Wang:pnas14}. However, the in-gap state observed here, especially induced by Ni adatoms, is different from previous ones. On the Ni adatom surface, the in-gap state locates at 1.9 eV, instead of 1.3 eV below $E_{\mathrm{F}}$. In addition, the band bends upward (see Figure\ \ref{Fig6}), which is opposite to the downward band bending observed in ref.\,\onlinecite{Wang:pnas14} and \onlinecite{Wang:jpcc13}. Furthermore, clear size dependent for the in-gap states are observed [See Figure\,\ref{Fig4}(b)], suggesting the in-gap states are originated from the deposited Ni on the surface.   

\begin {figure}
 \includegraphics [width=3.0 in,clip] {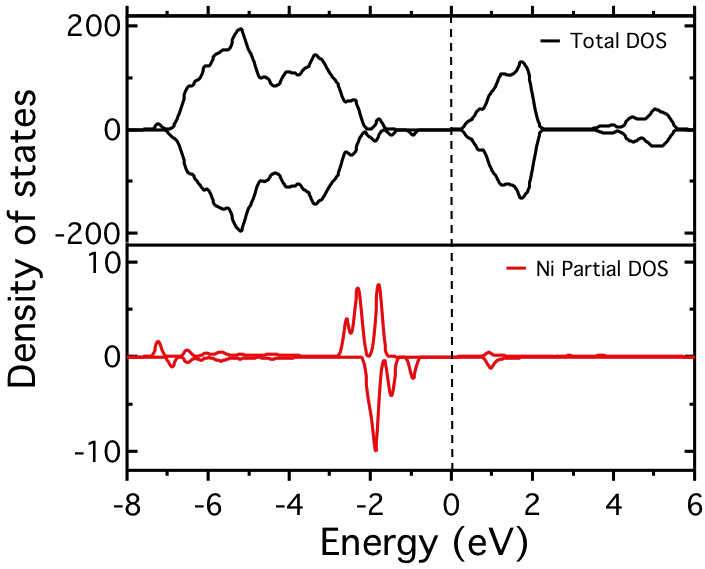}
 \caption{
(color online). PBE+U valence and conduction band density of states of the Ni$_{\mathrm{I}}$ adatom on the SrTiO$_3$(110)-(4~$\times$~1) surface. The upper and lower panel shows the total and Ni partial density of states, respectively.
}
\label{Fig5}
\end{figure}

\begin {figure}
 \includegraphics [width=3.0 in,clip] {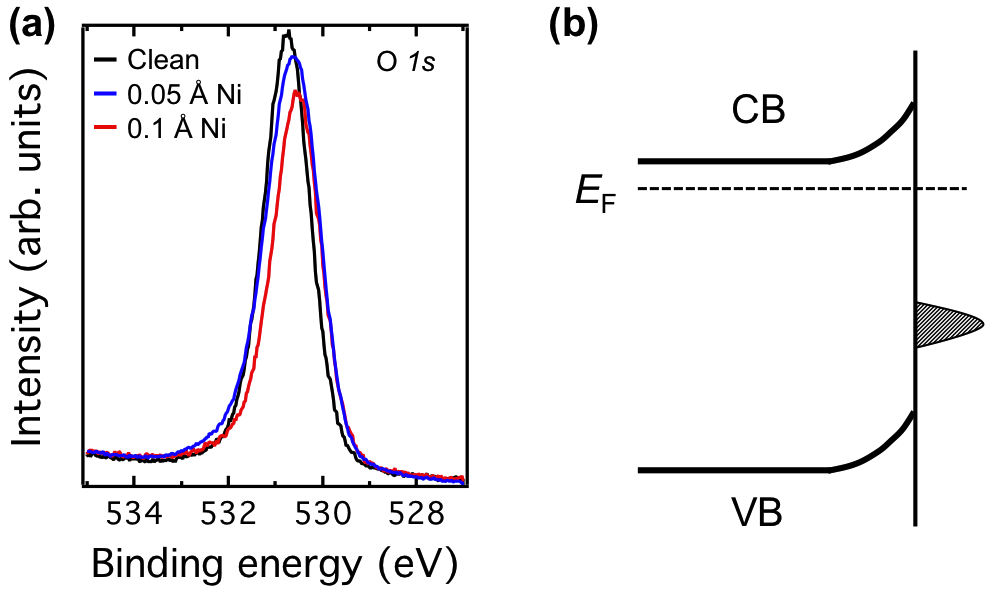}
 \caption{
(color online).  (a) Comparison of O $1s$ core-level photoemission spectra of  clean surface (black), and surfaces with 0.05\,{\AA} (blue), and 0.1\,{\AA} (red) Ni adatoms. The spectra are shifted 0.2 eV to lower binding energy after depositing Ni. All spectra were taken with the photon energy of 605 eV at RT. (b) Schematic diagram of the upward surface band bending induced by Ni adatom on the SrTiO$_3$(110)-(4~$\times$~1) surface.
}
\label{Fig6}
\end{figure}

To complement the photoemission spectra and obtain an understanding of the electronic properties of the Ni adatoms on the SrTiO$_3$(110)-(4~$\times$~1) surface, we have calculated the density of states for the most stable configuration of the Ni$_{\mathrm{I}}$ adatom (see Figure\ \ref{Fig5}). The most relevant feature is the appearance of in-gap states right above the valence band maximum, in line with the photoemission data. These states mainly originate from the Ni 3$d$ orbitals and locate on the surface. Since they are below the Fermi level (see the lower panel of Figure\ \ref{Fig5}), electrons from Nb dopant atoms in the SrTiO$_3$ layers can transfer into the surface states and an upward band bending occurs. This is consistent with the experimental results that valence band and O 1$s$ core-level spectra shift to lower binding energies after depositing Ni adatoms [see Figure\ \ref{Fig4}(a) and \ref{Fig6}(a)]. 

The charge state of single adatoms on oxide supports is important for their reactivity \cite{Lin:natchem11}, for example, both experimental and theoretical results have suggested that charged Au metal adatoms reduce the adsorption energy of small molecules as well as activation barrier for selected reactions \cite{Fang:pccp11}. The charge state of the Ni adatom is tentatively assigned to positive as the Ni 2$p$ core-level spectrum shifts to higher binding energy in Figure\,\ref{Fig4}(a), however, cautions are required due to the initial and final state effects in core-level photoemission spectroscopy \cite{Tao:ss08}. Further insight of the Ni charge state can be obtained from the Bader charge analysis on the basis of the DFT calculations.  Table\,\ref{Table1} lists the Bader charges for the Ni atom for the preferred configurations, and the corresponding magnetic moment within GGA+U scheme. It reveals that positively charged Ni adatoms are formed with Bader charges of +0.3 and +0.2 for the Ni$_{\mathrm{I}}$ and Ni$_{\mathrm{III}}$ adatom, respectively. Note that the Bader charge of the Ni adatoms obtained from the DFT calculations that do not consider the $n$-type doped SrTiO$_3$ samples were used in experiments. Correspondingly, the magnetic moment of 0.5 and 0.2 $\mu_\mathrm{B}$ is found for the Ni$_{\mathrm{I}}$ and Ni$_{\mathrm{III}}$ adatom, respectively. Apparently, the magnetic moment of the Ni atom is reduced but not completely quenched, compared with the magnetic moment of 0.6 $\mu_\mathrm{B}$ of bulk fcc Ni. 

\section{Conclusion}
In summary, we demonstrate that single Ni adatoms can be stabilized at the 2D porous titania on the SrTiO$_3$(110) surface at room temperature. Two types of Ni adatoms are formed by anchoring into the six- and ten-member nanopores, respectively. The Ni adatoms induce surface states at a binding energy of 1.9 eV and result in an upward band bending. Experimental and theoretical results suggest that Ni adatoms could be positively charged. Our study creates well-dispersed single adatom arrays on a well characterized oxide support, providing a model system to investigate single adatom catalytic and magnetic properties. \\

\acknowledgement

This work was supported by the Austrian Science Fund (FWF) under Project No. F45 and the ERC Advanced Research Grant ``OxideSurfaces''.




\begin{mcitethebibliography}{41}
\bibitem{note}
The adsorption energy, has
been evaluated with respect to the Nickel metal in fcc lattice/gas phase, is defined as
$E_{ads}(Ni)$= 1/2[$E_{tot}(Ni/slab)$-2$E_{Ni}$-$E_{tot}(slab)$].
Here $E_{tot}(Ni/slab)$ is the total energy of Ni-slab system is the equilibrium state, $E_{tot}(slab)$ the total energy of clean slab system, $E_{Ni}$
the energy of Ni in fcc lattice/gas phase. With this definition, more negative values reflect the strong interaction of Ni atom with the SrTiO$_3$ surface.
\mciteBstWouldAddEndPuncttrue
\mciteSetBstMidEndSepPunct{\mcitedefaultmidpunct}
{\mcitedefaultendpunct}{\mcitedefaultseppunct}\relax
\EndOfBibitem
\end{mcitethebibliography}

\providecommand*\mcitethebibliography{\thebibliography}
\csname @ifundefined\endcsname{endmcitethebibliography}
  {\let\endmcitethebibliography\endthebibliography}{}

\end{document}